\documentclass[twocolumn,showpacs,preprintnumbers,amsmath,amssymb,superscriptaddress,prb]{revtex4}
\usepackage{graphicx}
\begin{document}

\title{Mn-doped Ga(As,P) and (Al,Ga)As ferromagnetic semiconductors}

\author{J. Ma\v{s}ek}
\author{J. Kudrnovsk\'y}
\author{F. M\'aca}
\affiliation{Institute of Physics ASCR, Na Slovance 2, 182 21 Praha 8, Czech Republic}

\author{Jairo Sinova}
\affiliation{Department of Physics, Texas A\&M University, College
Station, TX 77843-4242, USA}
\author{A.H. MacDonald}
\affiliation{Department of Physics, University of Texas at Austin,
Austin TX 78712-1081, USA}

\author{R. P. Campion}
\author{B. L. Gallagher}
\affiliation{School of Physics and Astronomy, University of Nottingham, Nottingham NG7 2RD, UK}
\author{T. Jungwirth}
\affiliation{Institute of Physics ASCR, Cukrovarnick\'a 10, 162 53
Praha 6, Czech Republic}
\affiliation{School of Physics and Astronomy, University of Nottingham, Nottingham NG7 2RD, UK}

\begin{abstract}
A remarkable progress towards functional ferromagnetic semiconductor materials for spintronics has
been achieved in
p-type (Ga,Mn)As. Robust hole-mediated ferromagnetism has, however, been observed also in other
III-V hosts such as antimonides, GaP or (Al,Ga)As which opens a wide area of possibilities
for optimizing the host composition towards higher ferromagnetic Curie temperatures.
Here we explore theoretically ferromagnetism and 
Mn incorporation in Ga(As,P)  and (Al,Ga)As ternary hosts.
While
alloying (Ga,Mn)As with Al has only a small effect on the Curie
temperature we predict a sizable enhancement of Curie temperatures in the smaller lattice constant
Ga(As,P) hosts. Mn-doped Ga(As,P) is also favorable, as compared to (Al,Ga)As, 
with respect to the formation of  carrier and moment compensating
interstitial Mn impurities. In (Ga,Mn)(As,P) we find a marked decrease of the partial concentration of these detrimental 
impurities  with increasing P content. 
\end{abstract}
\maketitle

\section{Introduction}
Ordered by increasing band-gap and decreasing lattice constant of the III-V host, electronic states associated
with $\sim 1-10\%$ of substitutional Mn$_{\rm III}$ impurities may fall in the following three qualitative
categories:\cite{Jungwirth:2006_a,Schulthess:2005_a,Krstajic:2004_a} (i) The
main peak in the partial density of states of the majority-spin Mn $d^5$ electrons is well below the Fermi energy and
these states form a local moment close to 5 Bohr magnetons.
Mn acts here as an acceptor and the delocalized holes have a character
of the host states near the top of the valence band with a small admixture of the Mn $t_{2g}$ $d$-orbital weight.
In these metallic (III,Mn)V ferromagnets
the coupling between Mn local moments is mediated by delocalized band-holes via the
kinetic-exchange mechanism \cite{Dietl:1997_a,Matsukura:1998_a,Jungwirth:1999_a,Dietl:2000_a,Jain:2001_a}. (ii) The
second regime, often referred to as a double-exchange ferromagnet, is characterized by a stronger
hybridization of Mn $d$-states near the Fermi energy and by
holes occupying an impurity band which is detached from the host
semiconductor valence band. In this picture electrical conduction and
Mn-Mn exchange coupling are both realized through hopping within the impurity
band.\cite{Inoue:2000_a,Litvinov:2001_a,Durst:2002_a,Chudnovskiy:2002_a,Kaminski:2002_a,Berciu:2001_a,Bhatt:2002_a,Alvarez:2002_a,Xu:2005_c}
(iii) Finally, the substitutional   Mn$_{\rm III}$ impurities may undergo a  transition
from  a divalent ($d^5$) acceptor to a trivalent ($d^4$) neutral state. This strongly correlated
 $d^4$ center,
with four occupied  $d$-orbitals and a non-degenerate empty $t_{2g}$ $d$-level shifted deep into the host band gap,
may form as a result of a spontaneous (Jahn-Teller) lowering of the cubic symmetry
near the Mn site.  Systems with dominant
$d^4$ character of Mn impurities, reminiscent of a charge transfer insulator, will inevitably require
additional charge co-doping to provide for ferromagnetic coupling between dilute
Mn moments.\cite{Schulthess:2005_a,Krstajic:2004_a,Xu:2005_c}

\begin{figure}[h]

\hspace*{0cm}\includegraphics[width=.38\textwidth,angle=-90]{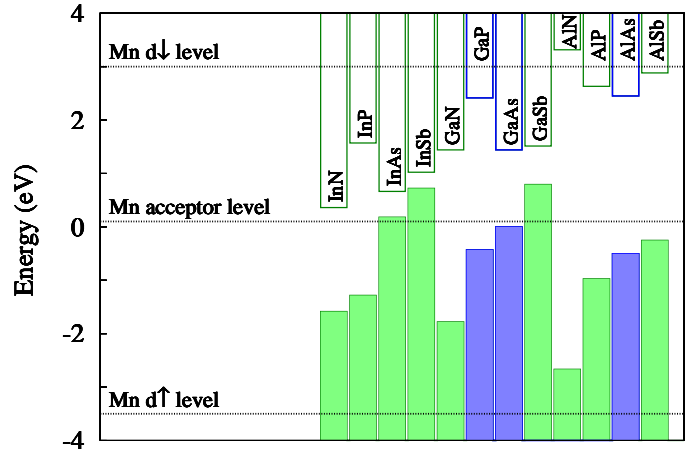}

\caption{Valence band and conduction band off-sets across the family of III-V semiconductors. Mn derived energy
levels are constant in this diagram according to the empirical internal reference rule.}
\label{ref_rule}
\end{figure}

The internal reference rule,\cite{Langer:1988_a,Dietl:2002_b}
which states that energy levels derived from Mn are constant across semiconductor compound families with properly aligned bands, serves as a useful guidance for associating individual III-V hosts with one of the three
categories listed above.  As seen from  Fig.~\ref{ref_rule},
Mn-doped InSb, InAs, and GaAs can be expected to exhibit long range Mn-Mn coupling
mediated by delocalized holes in the host valence band. This picture has indeed
been corroborated by a number of experimental studies.\cite{Jungwirth:2006_a} Measured Curie temperatures,
currently ranging from 7~K in the narrow gap, large lattice constant (In,Mn)Sb\cite{Csontos:2005_b} to 173~K in
the wider gap smaller, lattice constant (Ga,Mn)As,\cite{Wang:2004_c,Jungwirth:2005_b} are consistent with
the kinetic-exchange  model.

On the opposite side of the spectrum of III-V hosts, bulk (Ga,Mn)N  is an example
of a wide gap
material in which Mn does not provide for a significant hole doping.\cite{Graf:2002_a,Hwang:2005_b}
Considering the width of the band gap only, GaP
would fall
in the same category as GaN. As shown in Fig.~\ref{ref_rule}, however, the bands are significantly off-set to higher
energies in GaP and the resulting  Mn acceptor level
is shallow enough to lead  to a robust hole mediated
ferromagnetism at Mn dopings of several per cent. Experiments in (Ga,Mn)P reported to date
suggest\cite{Scarpulla:2005_a} the presence of an
impurity band in this III-V host, although more studies are needed to establish whether this character is intrinsic to
(Ga,Mn)P or occurs partly due to other unintentional impurities present in the
studied systems.\cite{Theodoropoulou:2002_a,Poddar:2005_a,Scarpulla:2005_a} (Note also that there is no sharp distinction between impurity band double-exchange and kinetic-exchange interactions; the former is simply a
strong coupling, narrow band limit of the latter.) A
factor of 2  smaller $T_c$ in (Ga,Mn)P, compared to (Ga,Mn)As prepared under similar growth
conditions,\cite{Scarpulla:2005_a}
indicates a marked suppression of the Curie temperature in this material due to the shorter range of
magnetic interactions in the hopping regime.

The theoretical work presented in this paper is based on  the currently best understood and highest $T_c$ (Ga,Mn)As
benchmark
material and explores whether Curie temperature can still be increased by introducing
elements from the higher rows in the periodic table. In particular, we focus on magnetic and structural
properties of Mn-doped Ga(As,P) and (Al,Ga)As ternary hosts. We exploit a
special circumstance that AlAs and GaAs lattice constants are very similar but GaP has a substantially smaller
lattice constant, and that AlAs and GaP have very similar band off-sets relative to
the smaller gap GaAs (see Fig.~\ref{ref_rule}). This allows us to disentangle
the roles in the effective magnetic interaction between Mn  and hole spins of the positions of Mn derived states
relative to the valence-band edge,
and of  the overlaps between anion (As or P) $p$-orbitals forming the top of the valence-band and the Mn $d$-orbitals.
Magnetic interactions are found to be more sensitive to the latter parameter. It explains why
alloying (Ga,Mn)As with Al has only a small effect on Curie
temperature\cite{Takamura:2002_a}
but we predict a sizable enhancement of the Curie temperature in the smaller lattice constant
Ga(As,P) ternary host. The mean-field $T_c$ calculations are presented in Section~\ref{magnet} together
with estimates of the range of magnetic interactions in different (Ga,Mn)(As,P) mixed crystals.

Structural properties of Mn-doped Ga(As,P) and (Al,Ga)As ferromagnetic semiconductors are studied in Section~\ref{formen}.
Under equilibrium growth conditions the incorporation of Mn ions into GaAs
is limited to less than 1\%. To circumvent the solubility problem a non-equilibrium, low-temperature
molecular-beam-epitaxy (LT-MBE) technique has to be used to achieve Mn doping concentrations
larger than 1\% at which ferromagnetism occurs. The highly doped LT-MBE (Ga,Mn)As systems show a strong
tendency to self-compensation by interstitial Mn$_{\rm I}$ impurities.\cite{Yu:2002_a,Maca:2002_a,Jungwirth:2005_b}
Correlations between acceptors (Mn$_{\rm Ga}$) and donors (Mn$_{\rm I}$) 
in GaAs\cite{Masek:2004_a,Jungwirth:2005_b}  are strong
due to the nearly covalent nature of bonding in the crystal. The cohesion energy of the covalent networks has a
maximum if the Fermi energy $E_F$  lies within the band gap. Whenever $E_F$  is shifted to the valence band
the strength of the bonds is reduced because of the occurrence of unfilled bonding states.
The tendency to self-compensation by Mn$_{\rm I}$ can therefore be expected to weaken in wider gap hosts
in which the substitutional Mn acceptor level is shifted deeper in the band gap. Another factor determining
the formation
energy of Mn$_{\rm I}$ is the size of the interstitial space in the lattice it fits in, i.e., the host
lattice constant. 
In Section~\ref{formen} 
we again use the comparison between Mn-doped Ga(As,P) and (Al,Ga)As hosts, with similar positions
of the substitutional Mn acceptor levels\cite{Bantien:1988_a,Clerjaud:1985_a} (see also Fig.~\ref{ref_rule})
but different lattice constants, to separate the roles of the two
factors. A significant decrease of the Mn$_{\rm I}$ partial concentration observed only
in Ga(As,P) indicates that the lattice parameter plays a more important role than the position of the 
 acceptor level in these mixed (III,Mn)V crystals.

We conclude our paper in Section~\ref{sum} with a brief discussion of the theoretical results  which we believe provide
motivation for a systematic experimental exploration of epitaxial (Ga,Mn)(As,P) ferromagnetic semiconductor
compounds.

\section{Magnetic interactions}
\label{magnet}
The discussion in this section is divided into three parts. First we assume the Schrieffer-Wolf transformation of the many-body
Anderson Hamiltonian and estimate trends in mean-field $T_c$ in (Ga,Mn)As, (Ga,Mn)P, and (Al,Mn)As
based on the effective interaction energy between local Mn moments and hole spins. In the second part we
calculate band structures  of Mn-doped Ga(As,P) and (Al,Ga)As 
ternary hosts, using the microscopic tight-binding coherent-potential
approximations (TBA/CPA),
and calculate corresponding mean-field Curie temperatures. Results of the {\em ab initio} LDA+U
calculations of the mean-field $T_c$ in (Ga,Mn)(As,P)  and of the range of  hole mediated Mn-Mn magnetic coupling are
presented in the last subsection. The latter
quantity is used to estimate the potential suppression of ferromagnetic interactions
due to stronger binding of the hole to Mn in the wider gap host.

\subsection{Qualitative estimates of the kinetic-exchange coupling and mean-field $T_c$}
Effects of strong Coulomb correlations in the Mn 3$d$-shell and hybridization with the host
semiconductor band states, which are at the heart of magnetism in these systems, can be qualitatively captured by
the Anderson Hamiltonian.\cite{Anderson:1961_a,Haldane:1976_a,Fleurov:1976_a} Here
Coulomb correlations are modelled by the on-site Hubbard potential which depends on the
occupation number of the $d$-states. The change in this effective potential when the number of occupied localized orbitals changes by one is parameterized by the Hubbard constant $U$.  The localized orbital part of the Anderson
Hamiltonian has an additional parameter, the Hund’s rule constant $J_H$. 
This parameter captures the local direct exchange physics which favors spin-polarized open shell atomic states. 
For the case of the Mn-$d^5$ conﬁguration, $J_H$ forces all  five  singly-occupied $d$-orbitals to 
align their spins in the ground state.

The Schrieffer-Wolff transformation  of the Anderson model removes the $p-d$ hybridization term in
the many body Hamiltonian and
leads to a description in which localized Mn $d$-states
interact with the valence band via a spin-spin interaction only.\cite{Schrieffer:1966_a}
Near the $\Gamma$-point, the strength of this  kinetic-exchange interaction can be parameterized by a constant
\begin{equation}
J_{pd}\propto \Omega_{u.c.}|V_{pd}|^2(1/|E_{d\uparrow}| +  1/|E_{d\downarrow}|) \; ,
\label{jpd-schrieffer}
\end{equation}
where
$\Omega_{u.c.}=a_{lc}^3/4$  is the volume of the unit cell of the zinc-blende crystal with a lattice
constant $a_{lc}$,
$V_{pd}\propto a_{lc}^{-7/2}$ represents the hybridization potential,\cite{Harrison:1980_a}
and $|E_{d\uparrow}|$ and  $|E_{d\downarrow}|$
are the distances of the occupied and empty atomic Mn
$d$-levels from $E_F$ ($|E_{d\downarrow}-E_{d\uparrow}|=U+5J_H$).\cite{Schrieffer:1966_a,Masek:1991_a,Timm:2004_b}

In the mean-field kinetic-exchange model the Curie temperature of
a III$_{1-x}$Mn$_x$V magnetic semiconductor scales as,\cite{Jungwirth:1999_a,Dietl:2000_a} $T_c\propto J_{pd}^2 x/\Omega_{u.c.}$, i.e.,
\begin{equation}
T_c\propto a_{lc}^{-11}(1/|E_{d\uparrow}| +  1/|E_{d\downarrow}|)^2 \; .
\label{tc-schrieffer}
\end{equation}
The effect of the second term in Eq.~\ref{tc-schrieffer} on $T_c$ in GaP or AlAs hosts, compared to GaAs,
can be estimated from Fig.~\ref{ref_rule}. Despite the twice as large band gap, the valence band off-set
is relatively small and the increase of $1/|E_{d\uparrow}|$ is almost completely compensated by the
decrease of $1/|E_{d\downarrow}|$. The second term in Eq.~\ref{tc-schrieffer} therefore leads to only a $4$\%
enhancement of the mean-field kinetic-exchange model $T_c$ in both GaP and AlAs hosts as compared to GaAs.

A much stronger enhancement of $T_c$ is obtained for GaP due to the first term in Eq.~\ref{tc-schrieffer}.
Considering $a_{lc}=5.653$~\AA\, for GaAs and $a_{lc}=5.450$~\AA\, for GaP,\cite{Vurgaftman:2001_a}
the mean-field kinetic-exchange model $T_c$ is by 50\% larger in (Ga,Mn)P than in (Ga,Mn)As.
For AlAs, on the other hand, no marked change in $T_c$ is expected from Eq.~\ref{tc-schrieffer} since
the material is nearly lattice matched with GaAs ($a_{lc}=5.661$~\AA, for AlAs).

\subsection{Microscopic tight-binding model calculations}
An effective single-particle TBA band-structure can be obtained
from the Anderson Hamiltonian by replacing the density operators
in the Hubbard terms with their mean values.\cite{Masek:1991_a} We
use this approach here, combined with the CPA, to calculate
microscopically mean-field Curie temperatures in Mn-doped
Ga(As,P) and (Al,Ga)As mixed crystals. Our TBA Hamiltonian includes the
8$\times$8 $sp^3$ term with second-neighbor-interaction integrals
describing the host semiconductor \cite{Talwar:1982_a} and terms
describing hybridization with Mn. Local changes of the crystal
potential at Mn are represented by shifted atomic levels.  The
parameters chosen for the atomic level shifts and  the hopping
amplitudes between atoms\cite{Talwar:1982_a,Masek:1991_a} provide
the correct band gap for the binary host   crystal and  the
appropriate exchange splitting of  the Mn $d$-states.  In particular, we considered in the TBA/CPA calculations
$U=3.5$~eV and $J_H=0.6$~eV, resulting in approximately 6~eV splitting of the peaks of the majority
and minority Mn $d$-projected density of states in (Ga,Mn)As. For the
ternary GaP$_y$As$_{1-y}$ and Al$_y$Ga$_{1-y}$As mixed crystals we
used the energy scale related to the GaAs band structure and
shifted the atomic level of the other components according to the
band off-sets of binary hosts shown in Fig.~\ref{ref_rule}. The
non-diagonal matrix elements of the TBA Hamiltonian appropriate
for the mixed crystals were obtained by linear interpolation
applied to the hopping integrals multiplied by the lattice
constant squared,\cite{Harrison:1980_a}
\begin{equation}
V(y)=\frac{yV(1)a_{lc}^2(1)+(1-y)V(0)a_{lc}^2(0)}{\left(ya_{lc}(1)+(1-y)a_{lc}(0)\right)^2}\;
\end{equation}
We used the same values
of $U$ and $J_H$ for all mixed crystals appealing to the nearly atomic-like, $d^5$ character of Mn impurity atoms.

Within the CPA,
disorder effects  associated with random distribution of Mn$_{\rm III}$ and of Al$_{\rm Ga}$ or
P$_{\rm As}$
appear in  the finite  spectral width  of  hole
quasiparticle states.
Since realizations with  near-neighbor Mn ions are
included within the disorder-averaged  TBA/CPA with the proper statistical
probability, short-range local moment  interactions (such as antiferromagnetic superexchange)
contribute to the final
magnetic  state. In uncompensated systems considered here, however, the long-range ferromagnetic
Mn-Mn coupling mediated
by holes dominates.

\begin{figure}[h]

\includegraphics[width=.32\textwidth,angle=-90]{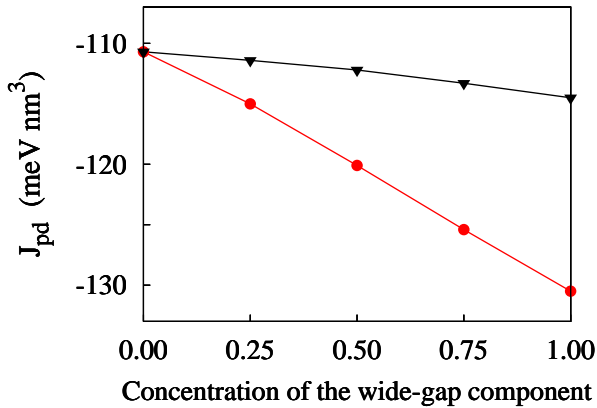}

\caption{TBA/CPA $J_{pd}$ in Al$_y$Ga$_{1-y}$As and GaP$_y$As$_{1-y}$ hosts doped with
$x=10$\% Mn.}
\label{jpd_tba}
\end{figure}

In Fig.~\ref{jpd_tba} we plot the dependence of the kinetic-exchange constant $J_{pd}$, derived from the splitting
of the TBA/CPA valence band at the $\Gamma$-point, as a function of the concentration of the wide-gap component
$y$.
As expected from the qualitative discussion above, $J_{pd}$ depends only weakly on the concentration of Al in the
Al$_y$Ga$_{1-y}$As host. A stronger variation is obtained for GaP$_y$As$_{1-y}$ with $J_{pd}$  enhancement
of 16\% at $y=1$, in good agreement with Eq.~\ref{jpd-schrieffer}.

\begin{figure}[h]

\includegraphics[width=.32\textwidth,angle=-90]{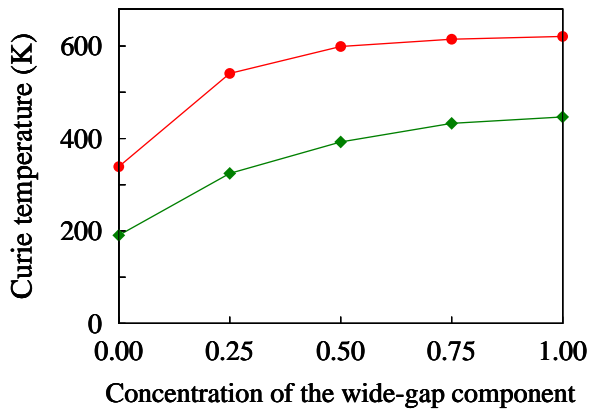}

\caption{TBA/CPA mean-field $T_c$ in  GaP$_y$As$_{1-y}$ hosts doped with
$x=5$ and 10\% Mn.}
\label{GaMNAsP_Tc}
\end{figure}

The TBA/CPA Curie temperatures are obtained  using  the
compatibility  of  the model with  the  Weiss  mean-field theory.
The strength of the Mn-Mn coupling  is characterized by the energy
cost, $E_{rev}$, of flipping one Mn moment, which can be calculated for a
given chemical composition.\cite{Masek:1991_a} 
$E_{rev}$ is proportional to the effective Weiss field
and to the mean-field Curie temperature, 
$k_BT_c=E_{rev}/6$.
In Fig.~\ref{GaMNAsP_Tc} we plot the TBA/CPA $T_c$ in
GaP$_y$As$_{1-y}$ host for 5\% and 10\% Mn concentrations. A
factor of 1.5-2 enhancement of $T_c$ for $y=1$ compared to the
GaAs host is again consistent with  qualitative predictions in
the previous section.

\subsection{Ab initio LDA+U theory}
In {\sl ab initio} approaches, the CPA description of 
disordered mixed crystals is combined with the local density approximation (LDA) to 
density-functional theory.
Electron correlations on the Mn $d$-shell are modelled
by including Hubbard interaction terms in the Hamiltonian
(LDA+U).\cite{Park:2000_a,Shick:2004_a,Wierzbowska:2004_a} 
The LDA+U/CPA method is implemented within the framework of the first-principles, tight-binding linear muffin-tin orbital 
approach.\cite{Turek:1997_a,Kudrnovsky:2004_a}
Hubbard parameters
used in the {\em ab initio} calculations were chosen to provide similar
exchange splitting of the Mn d-states as obtained in the TBA/CPA model calculations.

\begin{figure}[h]

\hspace*{-0.8cm}\includegraphics[width=.32\textwidth,angle=-90]{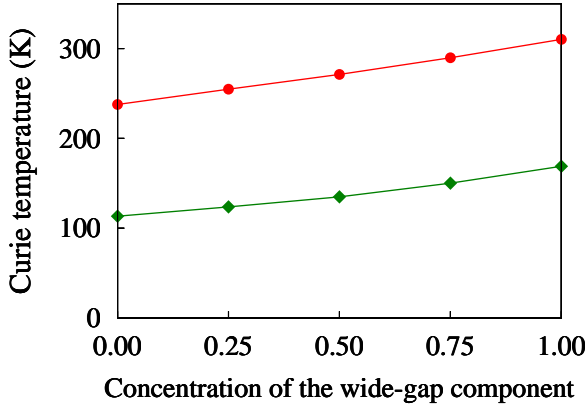}

\caption{{\em Ab initio} LDA+U mean-field $T_c$ in  GaP$_y$As$_{1-y}$ hosts doped with
$x=5$ and 10\% Mn.}
\label{tc_lmto}
\end{figure}

The mean field $T_{c}$'s, shown in  Fig.~\ref{tc_lmto},
are obtained again from the spin-flip energy $E_{rev}$ which is
calculated directly from the LDA+U/CPA
Green's functions.\cite{Liechtenstein:1987_a} 
$T_c$ clearly increases with increasing $y$ in the series of
Ga$_{1-x}$Mn$_{x}$As$_{1-y}$P$_{y}$ with fixed Mn doping $x$. 
Also in agreement with model calculations of previous sections, the LDA+U/CPA values of the mean-field $T_{c}$ are
proportional to the Mn concentration within the studied range of Mn dopings of 1-10\%.
Note that while the main Curie temperature trends with Mn and P doping are described consistently by the
different theoretical approaches employed in this paper the absolute values of $T_c$'s cannot be predicted with
a high quantitative accuracy. This is typical for microscopic theories of dilute moment ferromagnetic 
semiconductors.\cite{Jungwirth:2006_a} 
In the case of  TBA/CPA and LDA+U/CPA techniques we attribute the discrepancy to  significantly smaller
band gaps in the LDA+U/CPA spectra. For GaAs, e.g., the LDA+U band gap is 0.4~meV, compared to the TBA 
(and experimental) band gap
of 1.5~eV. The valence band edge in the LDA+U spectra is then shifted further from the majority Mn $d$-level and
the $p-d$ hybridization is suppressed, resulting in smaller mean-field $T_c$ values. 

\begin{figure}[h]

\vspace*{1cm}
\hspace*{-0.8cm}\includegraphics[width=.38\textwidth,angle=-90]{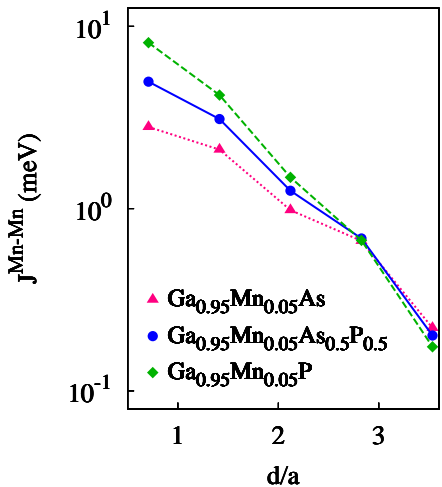}

\caption{{\em Ab initio} LDA+U range of Mn-Mn interactions in GaP$_y$As$_{1-y}$ hosts
with $y$=0, 0.5, and 1 along [110] crystal direction.}
\label{int_lmto}
\end{figure}

The energy $E_{rev}$ could also
be obtained from the interatomic exchange parameters $J_{ij}$
constructed by  mapping the LDA+U/CPA total energy on the Heisenberg
Hamiltonian:\cite{Liechtenstein:1987_a,Kudrnovsky:2004_a}
\begin{equation}
H = - \sum_{ij} J_{ij} \hat{e}_{i}\cdot \hat{e}_{j}\; ,
\end{equation}
where $\hat{e}_{i}$ is the local moment unit vector.
Since the 
moments induced on non-magnetic atom sites are small, the summation
can be restricted to sites occupied by Mn and
\begin{equation}
E_{rev} = 2 \sum_{j} J_{ij} \approx 2x \sum_{j}
J_{ij}^{Mn-Mn}.
\end{equation}
Individual $J_{ij}$ potentials characterize the spacial extent
of the Mn-Mn exchange coupling. This parameter is particularly useful 
for estimating corrections beyond mean-field 
theory in dilute moment systems. The mean-field approximation is more
reliable for sufficiently long-range character of carrier
mediated Mn-Mn coupling but tends to overestimate $T_c$ when the carriers become more localized and magnetic interactions
short-ranged.\cite{Sato:2004_a,Bergqvist:2004_a,Bouzerar:2004_a}  
As shown in Fig.~\ref{int_lmto}, the range of Mn-Mn exchange interactions is similar
within the whole family of  
Ga(As,P) ternary hosts and safely exceeds the average Mn-Mn moment distance. We therefore expect that 
results of 
mean-field theory reliably describe qualitative $T_c$ trends  
in Mn doped Ga(As,P) mixed crystals.

\section{Partial concentration of interstitial ${\rm\bf Mn}$}
\label{formen}

In the previous section we have considered all Mn impurities to substitute for the group-III element.
In  Mn-doped GaAs materials, however, a fraction of
Mn is incorporated during the growth at interstitial positions. These
donor impurities are likely to form pairs with substitutional Mn acceptors
in as-grown systems with approximately zero net moment,\cite{Blinowski:2003_a,Masek:2003_b,Edmonds:2004_a}
resulting in an effective free local-moment doping $x_{eff}=x_{s}-x_{i}$.
Here   $x_{s}$  and $x_{i}$ are partial  concentrations   of
substitutional and interstitial Mn, respectively. Although
Mn$_{\rm I}$ can be removed by low-temperature annealing, $x_{eff}$ will
remain smaller than the total nominal Mn doping, $x=x_s+x_i$. The substitutional Mn doping
efficiency is, therefore, one of the key parameters that may limit
$T_c$ also in  the wider gap III-V hosts.

\begin{figure}[h]

\hspace*{-0.8cm}\includegraphics[width=.32\textwidth,angle=-90]{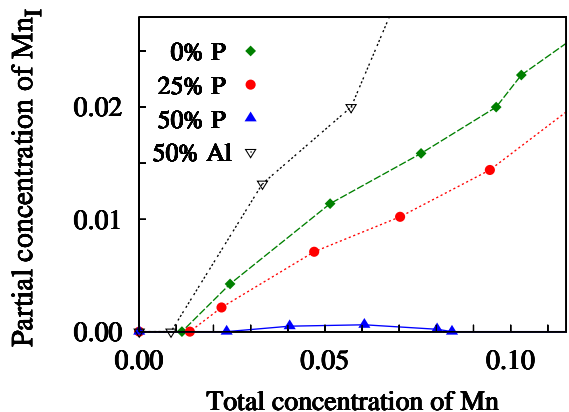}

\caption{Partial concentration of Mn$_{\rm I}$ as a function of total Mn concentration in
GaP$_y$As$_{1-y}$ 
with $y$=0, 0.25, and 0.5, and Al$_y$Ga$_{1-y}$As with $y$=0.5.}
\label{partial_conc}
\end{figure}

As discussed in the Introduction, the increasing ionicity and energy of the substitutional 
Mn acceptor level with increasing concentration of the wider gap element is one factor
that may reduce the tendency to self-compansation by Mn$_{\rm I}$. This mechanism should play a comparable
role in both Ga(As,P) and (Al,Ga)As hosts, given the similar experimental positions of the Mn acceptor level in 
AlAs, $E_a\approx$0.44~eV\cite{Bantien:1988_a},
and in GaP, $E_a\approx$0.4~eV.\cite{Clerjaud:1985_a} The formation energy of Mn$_{\rm I}$ may, however,
also increase due to a geometrical effect of a reduced size of the interstitial space in smaller lattice constant
hosts. Calculations for Ga(As,P) and for the nearly GaAs
lattice matched (Al,Ga)As, shown in Fig.~\ref{partial_conc}, indicate that the geometrical effect dominates.

The compositional dependence of the Mn$_{\rm I}$ partial concentration, plotted in Fig.~\ref{partial_conc},
is obtained from {\em ab initio}
formation energies of substitutional and interstitial Mn,
\cite{Masek:2002_a,Masek:2004_a,Jungwirth:2005_b} calculated for
the given GaP$_y$As$_{1-y}$ or Al$_y$Ga$_{1-y}$As host. For the smaller lattice constant Ga(As,P), the concentration of
Mn$_{\rm I}$ is significantly suppressed already at $y=0.25$ and
no Mn$_{\rm I}$ impurities are expected to form at $y>0.5$. If confirmed experimentally, 
this property might have a profound effect
on both structural and magnetic quality of  Mn-doped Ga(As,P) epilayers. In (Al,Ga)As, which has a slightly larger
lattice constant than GaAs, our results predict 
similar or even stronger tendency to Mn$_{\rm I}$ formation, as compared  to pure GaAs host.
\section{Discussion}
\label{sum}
The effect of alloying (Ga,Mn)As with P on Curie temperature has been discussed theoretically in a previous study by
Xu {\em et al.}\cite{Xu:2005_c} Assuming the double-exchange model for the entire family of Mn-doped Ga(As,P) hosts
the authors found no significant enhancement of $T_c$ with increasing P content. In {\em ab initio} theories the character
of magnetic interactions for a given (III,Mn)V dilute moment system shifts from the 
delocalized-hole kinetic-exchange towards
the impurity-band double-exchange when removing the Hubbard correlation potentials. Indeed, in
agreement with  Xu {\em et al.}, we
found only  weak dependence of the mean-field $T_c$ on the As/P ratio when replacing the LDA+U/CPA band structure
with energy spectra obtained using pure LDA/CPA. Extensive experimental studies of MBE-grown
(Ga,Mn)As favor the kinetic-exchange
model while data measured in the Mn ion-implanted GaP material are more readily interpreted within 
the double-exchange model.\cite{Jungwirth:2006_a} This may suggest that 
theoretical $T_c$ trends in Figs.~\ref{GaMNAsP_Tc} and \ref{tc_lmto} apply to Ga(As,P) hosts with lower P content
while the results of  Xu {\em et al.} (or the LDA/CPA theory) describe more reliably the opposite P-doping limit. 
The range of validity of either of the theoretical pictures can be ultimately  established only
by a detailed  study of epitaxial (Ga,Mn)(As,P) compounds with variable Mn and P concentrations and 
minimal number of other unintentional impurities or lattice defects.  

For practical reasons phosphorus cells are rarely mounted on laboratory III-V MBE systems. Note, however,
that useful complementary studies of hole localization effects on magnetic interactions in ternary hosts derived from GaAs
can be performed by alloying with the much more common Al. The tendency to localization and reduced
range of Mn-Mn exchange interactions should be comparable 
in (Al,Ga)As and Ga(As,P), given the similar position of the Mn acceptor level in these two 
hosts\cite{Bantien:1988_a,Clerjaud:1985_a} (see Fig.~\ref{ref_rule}). 
Experimental results\cite{Takamura:2002_a} in 
annealed, 5\% Mn-doped Al$_y$Ga$_{1-y}$As 
materials grown by LT-MBE with
$y$ ranging from 0 to 0.3 showed no marked dependence of $T_c$ on $y$. For the (Al,Ga)As ternary hosts which are lattice 
matched to GaAs and for which the kinetic-exchange picture implies $T_c$'s nearly independent of Al concentration, 
this experimental result can be interpreted as a signature of
comparably long-ranged Mn-Mn interactions in the studied (Al,Ga)As hosts as in the pure GaAs host. According 
to Figs.~\ref{ref_rule}-\ref{tc_lmto} it also implies, however, that $T_c$ should increase when alloying (Ga,Mn)As with P.

To conclude,
we found two motivating factors in the materials research of high temperature diluted ferromagnetic semiconductors
for performing a detailed experimental study of epitaxial 
(Ga,Mn)(As,P): We expect the Curie temperature to increase with increasing P concentration with an enhancement factor 
of up to $\sim$1.5 compared to (Ga,Mn)As with the same concentration of uncompensated Mn$_{\rm Ga}$
local moments. We also predict a significantly weaker tendency 
to carrier and moment self-compenstion by interstitial Mn$_{I}$ impurities in Ga(As,P) hosts compared
to pure GaAs. Finally we point out that Mn-doped (Al,Ga)As represents a useful complementary system to
(Ga,Mn)(As,P) for understanding trends in magnetic and structural properties of these wider gap ${\rm III-V}$ 
hosts.

We acknowledge fruitful discussions with  Tom Foxon, Andrew Rushforth, and Alexander Shick, 
and support from the Grant Agency of the Czech
Republic under Grant No. 202/05/0575 and  202/04/0583, from the Academy of Sciences of the
Czech Republic under Institutional Support No. AVOZ10100521 and AVOZ10100520, 
from the Ministry of Education of the
Czech Republic Center for Fundamental Research LC510 and COST P19 OC-150, from the UK
EPSRC under Grant No. GR/S81407/01, and from the National Science
Foundation under Grant No. PHY99-07949.


\end{document}